\documentclass[12pt,preprint]{aastex}
\usepackage{amssymb}
\usepackage{graphicx}
\usepackage{longtable}
\usepackage{amsmath}

\begin{document}

\title{Long GRBs are metallicity-biased tracers of star formation: evidence from host galaxies and redshift distribution}
\author{F. Y. Wang$^{1,2}$ and Z. G. Dai$^{1,2}$}

\affil{$^1$School of Astronomy and Space Science, Nanjing
University,
Nanjing 210093, China \\
$^2$Key Laboratory of Modern Astronomy and Astrophysics (Nanjing
University), Ministry of Education, Nanjing 210093, China}

\altaffiltext{}{E-mail: fayinwang@nju.edu.cn (FYW); dzg@nju.edu.cn
(ZGD)}

\begin{abstract}
We investigate the mass distribution of long gamma-ray burst (GRB)
host galaxies and the redshift distribution of long GRBs by
considering that long GRBs occur in low-metallicity environments. We
calculate the upper limit on the stellar mass of a galaxy which can
produce long GRBs by utilizing the mass-metallicity (M-Z) relation
of galaxies. After comparing with the observed GRB host galaxies
masses, we find that the observed GRB host galaxy masses can fit the
predicted masses well if GRBs occur in low-metallicity
$12+\log\rm(O/H)_{\rm KK04}<8.7$. GRB host galaxies have low
metallicity, low mass, and high star formation rate compared with
galaxies of seventh data release of the Sloan Digital Sky Survey. We
also study the cumulative redshift distribution of the latest
\emph{Swift} long GRBs by adding dark GRBs and 10 new GRBs redshifts
from TOUGH survey. The observed discrepancy between the GRB rate and
the star formation history can be reconciled by considering that
GRBs tend to occur in low-metallicity galaxies with
$12+\log\rm(O/H)_{\rm KK04}<8.7$. We conclude that the metallicity
cutoff that can produce long GRBs is about $12+\log\rm(O/H)_{\rm
KK04}<8.7$ from the host mass distribution and redshift
distribution.
\end{abstract}


\section{Introduction}
Gamma-ray bursts (GRBs) are the most luminous explosions in the
Universe. Due to their high luminosities, GRBs can be observed
throughout most of the observable Universe (Lamb \& Reichart 2000;
Ciardi \& Loeb 2000; Cucchiara et al. 2011). So GRBs are ideal tools
to study early universe properties, including star formation rate,
reionization and metal enrichment history (Wang \& Dai 2009; Wang et
al. 2012), and so on. According to the duration, GRBs are usually
classified into two classes: long GRBs and short GRBs (Kouveliotou
et al. 1993). For long GRBs, their host galaxies are typically
irregular galaxies with high star formation rate and, especially, a
small fraction of long GRBs are associated with Type Ib/c supernovae
(SNe) (for a review, see Woosley \& Bloom 2006). These GRBs are
nearby and sub-luminous. The well-studied SNe that accompany GRBs
show evidence for broad lines, indicative of high-velocity ejecta.
This type of SNe is a subclassification of Type Ic SNe, called Type
Ic-BL. In contrast, short GRBs are usually found at nearby
early-type galaxies, with little star formation (for a review, see
Nakar 2007). But some short GRB hosts,  such as GRBs 100625A and
101219A, are early-type galaxies with moderate star formation
(Berger 2009; Fong et al. 2013). So studies on the host galaxies
properties are crucial to understand the progenitors and central
engines of GRBs.

Some theoretical studies of long GRB progenitors using stellar
evolution models suggest that low metallicity may be a necessary
condition for a long GRB to occur. For popular collapse models of
long GRBs, stars with masses $>30M_\odot$ can be able to create a
black hole (BH) remnant (Woosley et al. 1993; Hirschi, Meynet \&
Maeder 2005). The preservation of high angular momentum and
high-stellar mass at the time of collapse (Woosley 1993; MacFadyen
\& Woosley 1999) is crucial for producing the relativistic jet and
high luminosity. Low-metallicity ($0.1-0.3Z_\odot$) progenitors can
theoretically retain more of their mass due to smaller line-driven
stellar winds (Kudritzki \& Puls 2000; Vink \& de Koter 2005), and
hence preserve their angular momentum (Yoon \& Langer 2005; Yoon,
Langer \& Norman 2006; Woosley \& Heger 2006). Because the
wind-driven mass loss of massive stars is proportional to the
metallicity.

Observations of long GRB host galaxies also show that they are
typically in low metallicity environment, for several local long GRB
host galaxies (Sollerman et al. 2005; Stanek et al. 2006) as well as
in distant long GRB hosts (Fynbo et al. 2003; Christensen et al.
2004; Gorosabel et al. 2005; Fruchter et al. 2006; Prochaska et al.
2007). Stanek et al. (2006) found that the host galaxy metallicities
of five $z<0.25$ long GRBs were lower than equally luminous dwarf
irregular galaxies and concluded that the upper metallicity limit
for producing GRBs is about $0.15~Z_\odot$. Modjaz et al. (2008)
found that long GRB host galaxies had lower metallicities than the
host galaxies of nearby broad-lined Type Ic supernovae. Savaglio et
al. (2009) examined 46 long GRB host galaxies with 17 metallicity
measurements, and found that the hosts had subsolar metallicity with
an average metallicity of 1/6 solar value for 17 of the hosts.

Using a small sample of 5 host galaxies, Han et al (2010) found that
the metallicities of the host galaxies tended to fall below the low
redshift mass-metallicity (M-Z) relation defined by Sloan Digital
Sky Survey (SDSS) catalog. Likewise, Levesque et al. (2010a)
compared a much broader sample of long GRB host galaxies and found a
similar offset, with long GRB host galaxies exhibiting lower
metallicities compared to SDSS galaxies of similar masses. But a few
GRB hosts with high metallicity are observed, so that the role of
metallicity in driving the GRB phenomena remains unclear and it is
still debated (Price et al. 2007; Wolf \& Podsiadlowski 2007;
Kocevski, West \& Modjaz 2009; Graham et al. 2009; Svensson et al.
2010). For excellent reviews, see Fynbo et al. (2013) and Levesque
(2013). Mannucci et al. (2011) proposed that the apparent long GRB
preference for low-metallicity hosts is due to the fundamental
metallicity relation (Kocevski \& West 2011). Long GRB hosts have
low metallicity because they are effectively selected based on the
intense star-formation. By comparing the metallicity of the GRB
hosts, Type Ic (Ic-bl) supernovae (SNe), and Type II SNe to each
other and to the metallicity distribution of star-forming galaxies,
Graham \& Fruchter (2013) found that low metallicity must be a
fundamental property of long GRB hosts, rather due to the
fundamental metallicity relation. Trenti et al. (2013) found that
the empirical relation between GRB rate and SFR is due to the GRB
preference for low-metallicity.

In this paper, we investigate whether long GRBs occur in
low-metallicity environments from observations. We first use
empirical models based on the measurements of the redshift evolution
of the M-Z relation to estimate the upper limit of the stellar mass
of a galaxy that can efficiently produce a GRB, and test the
suggestion that GRBs preferentially form in low-metallicity
environments with observations. We also compare the GRB host
properties with SDSS observations. Then we present the latest
\emph{Swift} GRB cumulative redshift distribution considering GRBs
occur in low-metallicity environment. The structure of this paper is
as follows: in section 2, we give the theoretical mass distribution
of GRB host galaxy and compare this with observations. In section 3,
we fit the cumulative redshift distribution of \emph{Swift} long
GRBs by considering that GRBs occur in low-metallicity environments.
Conclusions and discussions are shown in section 4.

\section{The mass distribution of GRB host galaxy}
In this section, we first estimate the upper limit of the galaxy
mass that is capable of producing a GRB using the M-Z relation. Then
we compare this result with the host galaxy masses from
observations. Last, we will compare the host galaxy properties with
galaxies observed by SDSS.

\subsection{Theoretical model}
\begin{figure}\centering
\includegraphics[width=0.5\textwidth]{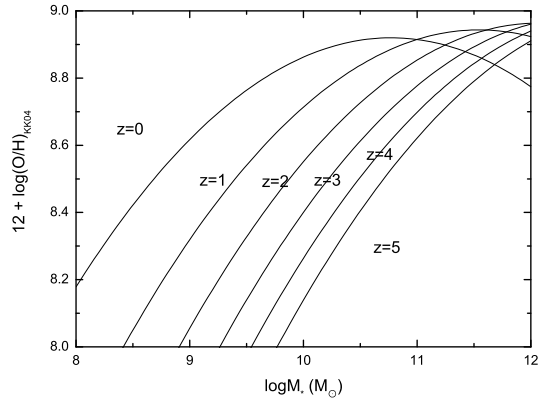} \caption{Redshift evolution of
galaxy mass-metallicity relation derived by Savaglio et al. (2005).}
\label{MZ}
\end{figure}

Using $\sim$53,000 galaxies from the SDSS, Tremonti et al. (2004)
found a tight correlation between stellar mass and metallicity, so
called mass-metallicity (M-Z) relation, which was discovered by
Lequeux et al. (1979) for the first time. Savaglio et al. (2005)
investigated the empirical redshift-dependent M-Z relation using
galaxies at $0.4 < z < 2.0$ from the Gemini Deep Deep Survey (GDDS)
and Canada-France Redshift Survey (CFRS). The redshift-dependent M-Z
relation can be parameterized as (Savaglio et al. 2005)
\begin{eqnarray} \label{eq:MZRz}
12+\log\rm(O/H)_{\rm KK04} &=& -7.59+2.53\log M_{\star}-0.097\log^2M_{\star} \nonumber \\
 &&+5.17\log t_{\rm H}-0.39\log^2t_{\rm H}\nonumber \\
 && -0.43\log t_{\rm H}\log M_{\star},
\end{eqnarray}
where KK04 represents the metallicity scale of Kobulnicky \& Kewley
(2004), $t_{\rm H}$ is the Hubble time at redshift $z$ in Gyr and
$M_{\star}$ is the galactic stellar mass in unit of solar mass. The
Hubble time at redshift $z$ is given by
\begin{equation}
t_H(z)=\frac{1}{H_0}\int_z^{\infty}\frac{dz'}{(1+z')\sqrt{\Omega_m(1+z)^3+\Omega_\Lambda}}.
\end{equation}
In the whole paper, we use $\Omega_m=0.27$, $\Omega_\Lambda=0.73$
and $H_0$=71 km~s$^{-1}$~Mpc$^{-1}$ from the \emph{Wilkinson
Microwave Anisotropy Probe} (WMAP) seven-year data (Komatsu et al.
2011). Figure \ref{MZ} shows the metallicity as a function of
stellar mass at different redshifts. Studies at the redshift range
$z\sim 1-3$ (Mannucci et al. 2009; Zahid et al. 2011; Finkelstein et
al. 2011), and high redshift $z\sim 3-5$ (Laskar et al. 2011), show
that the M-Z relation may keep the same overall trend.

\begin{figure}\centering
\includegraphics[width=0.5\textwidth]{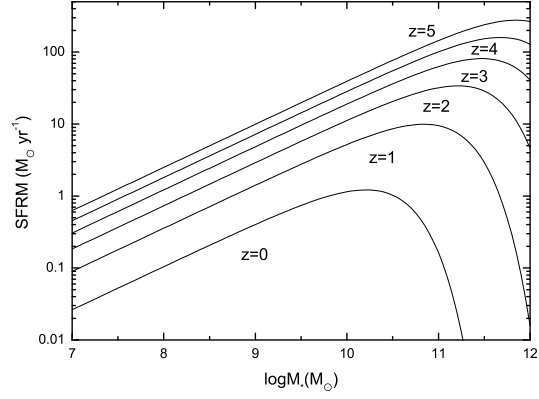} \caption{Average star formation rate as a function of stellar mass
at different redshifts. The peak rates evolve to high stellar mass
with increasing redshift.} \label{SFRM}
\end{figure}

In order to model the effects of a metallicity cutoff on the mass
distribution of GRB host galaxies, the number density of galaxies
and the number of stars being produced in those galaxies as a
function of galactic stellar mass are also needed. Drory \& Alvarez
(2008) found the star-formation rate as a function of stellar mass
and redshift spanning $9< \log M_* < 12$ and $0 < z < 5$ using data
from FORS Deep Field survey. They parameterize the observed star
formation rate-stellar mass relation as
\begin{equation}\label{eq:SFR}
\mathrm{SFRM} (M_{\star},z) = \mathrm{SFRM_0}
\left(\frac{M_{\star}}{M_0}\right)^{\beta}{\rm
exp}\left(-\frac{M_{\star}}{M_0}\right),
\end{equation}
where $\beta=0.6$, $\rm SFRM_0$ is the overall normalization and
$M_0$ represents the break mass the mass above which the average
star formation rate begins to decline. The best fit
parameterizations from Drory \& Alvarez (2008) are
\begin{eqnarray} \label{eq:SFRz}
\mathrm{SFRM_0}(z) \approx  3.01(1+z)^{3.03}M_\odot\mathrm{~yr^{-1}}, &  \\
M_0(z) \approx 2.7\times10^{10}(1+z)^{2.1}M_\odot.
\end{eqnarray}
We show star-formation rate as a function of stellar mass from $z=0$
to $z=5$ in Figure \ref{SFRM}.

The galactic stellar mass function is commonly described by a
Schechter function. We use the Schechter form derived by Fontana et
al. (2006) measured from the GOODS-MUSIC field,
\begin{equation}
\phi(M, z) =
\phi^{*}\left(z\right)\ln\left(10\right)[10^{M-M^{*}\left(z\right)}]^{1+\alpha^{*}\left(z\right)}\exp(-10^{M-M^{*}\left(z\right)}),
\label{eq:Mz}
\end{equation}
where $M=\log_{10}\left(M_{*}/M_{\sun}\right)$, $M_{*}$ is the
stellar mass of the galaxy and the parametric functions obey:
\[
\phi^{*}\left(z\right) = \phi^{*}_{0}\left(1+z\right)^{\phi_{1}^{*}}
\]
\[
\alpha^{*}\left(z\right) = \alpha_{0}^{*}+\alpha_{1}^{*}z
\]
\[
M^{*}\left(z\right) = M_{0}^{*} + M_{1}^{*} z + M_{2}^{*} z^{2}
\]
The parameter values are given as: $\phi^{*}_{0}=0.0035$,
$\phi^{*}_{1}=-2.20$, $\alpha_{0}^{*}=-1.18$,
$\alpha_{1}^{*}=-0.0082$, $M_{0}^{*}=11.16$, $M_{1}^{*}=0.17$ and
$M_{2}^{*}=-0.07$.

\begin{figure}\centering
\includegraphics[width=0.5\textwidth]{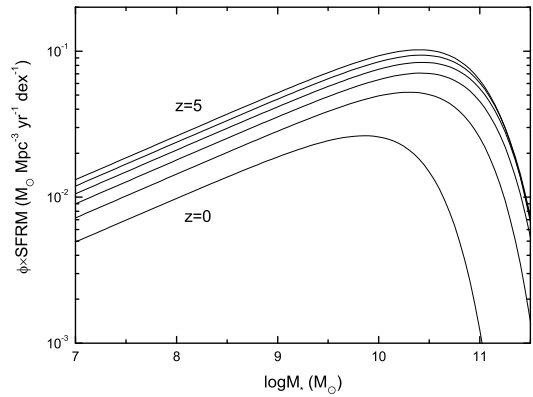} \caption{Total star formation rate as a function of stellar mass
at different redshifts ($z=0$ to $z=5$ from bottom to top).}
\label{tSFR}
\end{figure}

In order to know the total number of stars being produced as a
function of stellar mass, we compute the galaxy-weighted star
formation rate by multiplying equation (\ref{eq:SFR}) by equation
(\ref{eq:Mz}). The total star formation rate is shown in Figure
\ref{tSFR}. The weighted star formation rate peaks at intermediate
masses between $10^{10}M_\odot$ and $10^{11}M_\odot$.

\subsection{Comparison to GRB host galaxy observations}

\begin{figure}\centering
\includegraphics[width=0.5\textwidth]{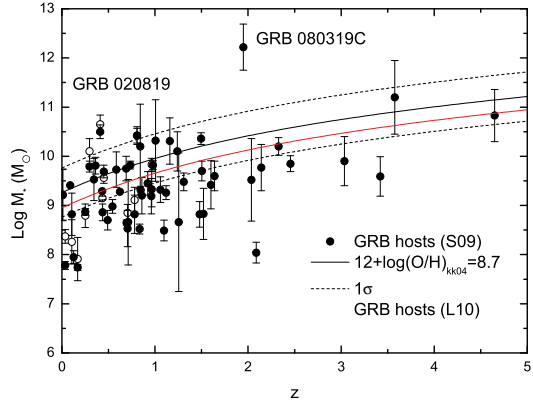} \caption{GRB host galaxy mass distribution.
The host galaxy masses are taken from Savaglio et al. (2009) and
GHostS. The solid lines represent the upper limits of the stellar
mass of a GRB host galaxy given a metallicity cutoff of
$12+\log\rm(O/H)_{\rm KK04}=8.7$ (black), and $12+\log\rm(O/H)_{\rm
KK04}=8.6$ (red). The dashed lines represent the scatter in the
upper limit imposed by the 1$\sigma$ scatter of the M-Z relation.}
\label{hostmass}
\end{figure}

The GRB host galaxy can be observed when the bright emission of the
GRB is gone. Often the ionized gas (the H~II regions) is emitting
with sufficient intensity so emission lines can be detected
(Savaglio 2006). Savaglio et al. (2009) compiled a sample of 40 long
GRB host galaxies through a combination of optical and NIR
observations. They found that GRB host galaxies exhibit a wide range
of stellar mass and star formation rates, although as a whole they
tend toward low stellar mass, relatively dim, high specific
star-forming systems. They also estimated that the average stellar
mass is $10^{9.3}M_\odot$. Besides the data of these 40 host
galaxies, we also use the latest GRB host data from GRB Host Studies
(GHostS), which is accessible at the Web site www.grbhosts.org. The
total number of GRB host galaxies with estimated mass is 58
(hereafter SG sample), which are listed in Table 1. The
metallicities are converted to KK04 scale based on Table 3 of Kewley
\& Ellison (2008). In Figure \ref{hostmass}, we show the host galaxy
mass distribution of SG sample with black dots. The solid line
represents the upper limits of the stellar mass of a GRB host galaxy
given a metallicity cutoff of $12+\log\rm(O/H)_{\rm KK04}=8.7$. The
dashed lines represent the scatter in the upper limit imposed by the
1$\sigma$ scatter of the M-Z relation. The most long GRB host galaxy
masses are well below upper limits of the stellar mass of a GRB host
galaxy given a metallicity cutoff of $12+\log\rm(O/H)_{\rm
KK04}=8.7$ excluding a few GRBs. Levesque et al. (2010a, hereafter
L10) presented a sample of 16 long GRB host galaxies, including 13
GRBs with host galaxy mass determination. They also concluded that
long GRBs tend to occur in host galaxies with lower metallicities
than the general population. We show the host mass distribution in
Figure \ref{hostmass} with open dots. The long GRB host galaxy
masses of L10 are below the upper limits due to metallicity cutoff
of $12+\log\rm(O/H)_{\rm KK04}=8.7$ except for a few GRBs. There are
some uncertainties when measuring the metallicities of GRBs'
explosion regions at high-redshifts, such as chemical inhomogeneity
(Levesque et al. 2010b; Niino 2011). Long GRB locates in a lower
metallicity region of its host, but the galactic average measured.
We will discuss the probability as follows. For GRB 020819, the host
metallicity is $12+\log\rm(O/H)_{\rm KK04}=9.0\pm0.1$ (Levesque et
al. 2010b), which is the highest metallicity determined for a long
GRB to date. Interestingly, some theoretical models predict that
long GRBs can occur in high metallicity environment (Podsiadlowski
et al. 2010; Georgy et al. 2012). For GRB 080319C, the apparent
brightness of the host galaxy may be overestimated. This burst is a
dark burst with large extinction ($A_V=0.65$ mag) (Perley et al.
2009). There is an intervening Mg II system at $z=0.81$ along the
line of sight of GRB 080319C (Fynbo et al. 2009). The apparent
brightness of the host galaxy may be overestimated due to the
contamination of Mg II system. So the mass of host galaxy should be
overestimated.

\begin{figure}\centering
\includegraphics[width=0.5\textwidth]{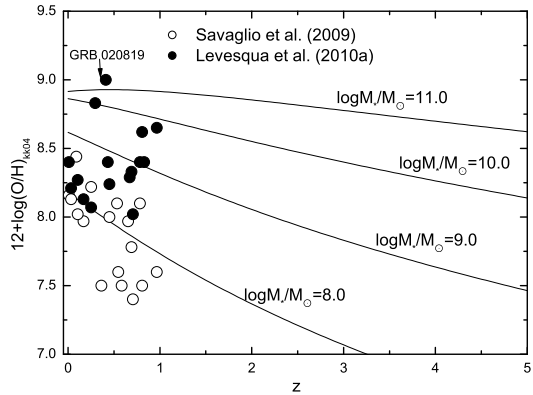} \caption{Metallicity as a
function of redshift. The observed GRB host galaxy metallicities are
taken from Savaglio et al. (2009) and Levesque et al. (2010a). The
curves are predictions from the empirical model of equation
(\ref{eq:MZRz}), for different stellar masses.} \label{SGmetal}
\end{figure}

In Figure \ref{SGmetal}, we show the observed metallicities of GRB
host galaxies taking from Savaglio et al. (2009) and Levesque et al.
(2010a). The solid lines represent the predictions from the
empirical model of equation (1), for different stellar masses. We
can see that the long GRBs are below the theoretical prediction of
equation (1) except for GRB 020819. As discussed above, more precise
localization of this burst is needed to draw robust conclusions. The
recent observations of the afterglow of GRB 090323 at redshift
$z=3.57$ by Savaglio et al. (2012) shows evidence for two damped
Ly$\alpha$ (DLA) systems with supersolar metallicities. But
high-resolution observation is needed to check whether one of them
is the host galaxy.

\begin{figure}\centering
\includegraphics[width=0.5\textwidth]{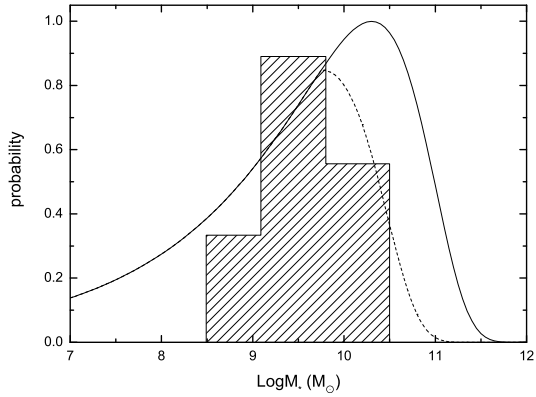} \caption{GRB host mass distribution as measured by sub-SG sampple
between $0.75 < z < 1.25$. The solid line is the galaxy-weighted
star formation rate as a function of galactic stellar mass at $z=1$.
The mass limit due to sharp metallicity cutoff of
$12+\log\rm(O/H)_{\rm KK04}=8.7$ is represented by the dashed line.
The peak of the sub-SG sample is much more consistent with the the
expected peak of a biased galaxy-weighted star formation rate.}
\label{probability}
\end{figure}

If long GRBs are unbiased tracers of star formation throughout the
universe, their observed host mass distribution should have a
largest probability at the peak in the galaxy-weighted star
formation rate. We will test this expectation below. The long GRB
host galaxy median mass of SG sample is $10^{9.45}M_\odot$, which is
obviously contradict the prediction from Figure \ref{tSFR}. We
should compare the observed host mass distribution to the expected
mass distribution of all star-forming galaxies at a given redshift,
because the detection effects should not affect this region of the
observed distribution. This problem is also discussed in Kocevski et
al. (2009). In order to use much more observation data, we carry out
this test around $z=1$. In Figure \ref{probability}, we show the GRB
host mass distribution of SG sample between $0.75\leq z \leq 1.25$
as histogram. The number of long GRBs with host mass determination
in this redshift range is 16. The galaxy-weighted star formation
rate as a function of galactic stellar mass at $z=1$ is shown as the
solid line in Figure \ref{probability}. After considering the mass
limits due to sharp metallicity cutoff of $12+\log\rm(O/H)_{\rm
KK04}=8.7$, we calculated the galaxy-weighted star formation rate as
a function of galactic stellar mass, which is shown as the dashed
line in Figure \ref{probability}. The average host galaxy mass of
this sub-SG sample is $10^{9.50}M_\odot$. After considering the
metallicity cutoff, the peak of the galaxy-weighted star formation
rate will shift from $10^{10.3}M_\odot$ to $10^{9.7}M_\odot$, which
is much more consistent with the observation.

\subsection{GRB host galaxy observations compared to SDSS galaxy}
\begin{figure}
\centering
\includegraphics[width=0.5\textwidth]{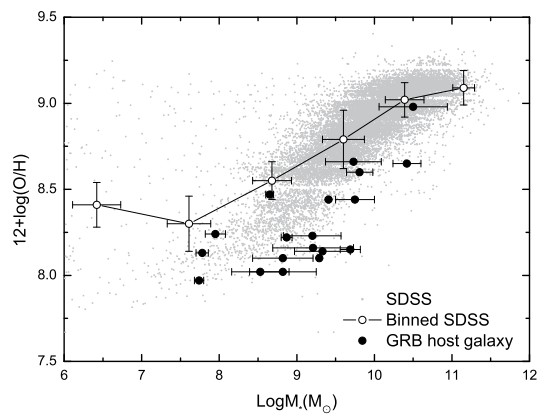} \caption{Comparsion between the
meatllicities of GRB host galaxies and SDSS galaxies. The gray
points represent individual galaxies of SDSS. The open dots are SDSS
binned metallicities, and filled dots are the metallicities of GRB
host galaxies from observations. The metallicities of GRB hosts are
well below the values of SDSS galaxies.} \label{metalmass}
\end{figure}

\begin{figure}
\centering
\includegraphics[width=0.5\textwidth]{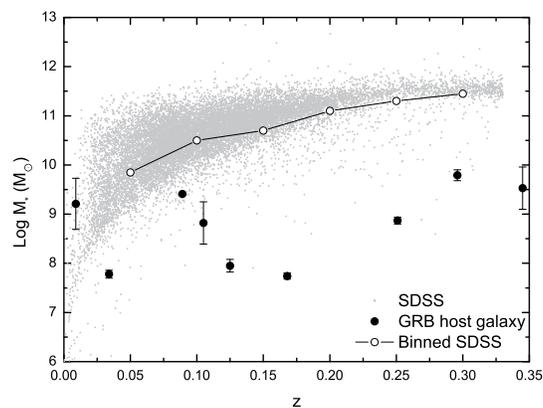} \caption{Comparsion between the
masses of GRB host galaxies and SDSS galaxies. The gray points
represent individual galaxies of SDSS. The filled dots are the
masses of GRB host galaxies from observations, and the open dots are
binned SDSS masses. } \label{redmass}
\end{figure}

\begin{figure}
\centering
\includegraphics[width=0.5\textwidth]{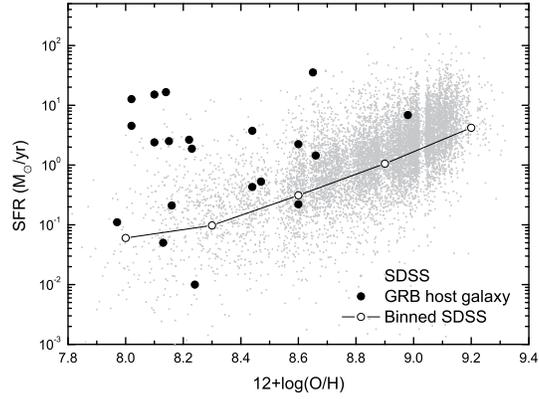} \caption{Comparsion between the
metallicities and SFRs of GRB host galaxies and SDSS galaxies. The
gray points represent individual galaxies of SDSS. Filled dots are
the value of GRB host galaxies from observations, and the open dots
are SDSS binned SFRs.} \label{metalsfr}
\end{figure}

\begin{figure}
\centering
\includegraphics[width=0.5\textwidth]{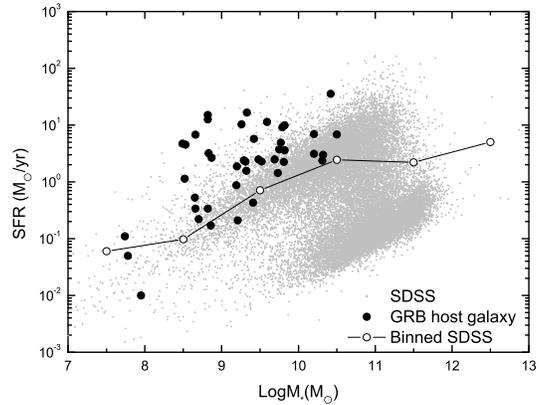} \caption{Comparsion between the
SFR of GRB host galaxies and SDSS galaxies. The gray points
represent individual galaxies of SDSS. Filled dots are the SFRs of
GRB host galaxies from observations and open dots are binned SDSS
SFRs.} \label{sfrsdss}
\end{figure}

\begin{figure}
\centering
\includegraphics[width=0.5\textwidth]{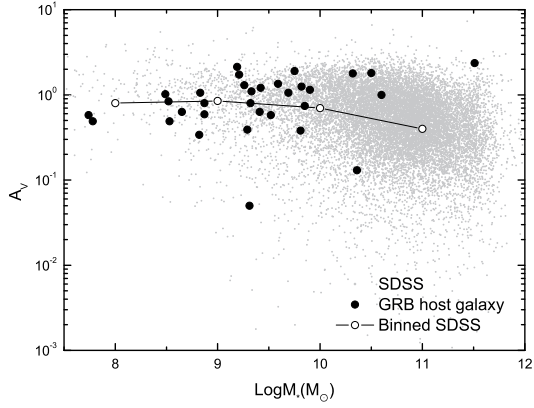} \caption{Comparsion between the
dust extinction of GRB host galaxies and SDSS galaxies. The gray
points represent galaxies observed by SDSS. Filled dots are the dust
extinction of GRB host galaxies from observations and open dots are
binned SDSS dust extinction $A_V$.} \label{avsdss}
\end{figure}

We use the galaxies that are well measured by SDSS-DR7 project
(Abazajian et al. 2009). The derived galaxy properties from the
MPA-JHU are available at http://www.mpa-garching.mpg.de/SDSS/DR7,
which contains 927 552 galaxies. The detailed analysis process can
be found in Kauffmann et al. (2003) and Salim et al. (2007) for
stellar masses, Brinchmann et al. (2004) for SFRs and Tremonti et
al. (2004) for metallicities. We use the information of galaxies,
including dust extinction $A_V$, stellar masses, star formation
rates, and metallicities. Although the redshifts of galaxies from
SDSS are less than 0.6, the redshifts of our GRB host galaxy sample
are also mainly less than 1.0. The median redshift of SDSS is 0.15,
and the median redshift of GRB host is 1.13. In
Figure~\ref{metalmass}, we compare the metallicities of GRB host
galaxy and SDSS galaxy with same masses. The gray points represent
individual galaxies of SDSS. The open dots are binned metallicities
of SDSS galaxies in mass range, and filled dots are the
metallicities of GRB host galaxy from observation. The metallicities
of GRB host galaxy are well below those of SDSS galaxies. So GRBs
prefer to occur in low-metallicity galaxies. Figure~\ref{redmass}
illustrates the GRB host galaxy masses compared to masses of SDSS
galaxies at the same redshift. The gray points represent individual
galaxies of SDSS. The filled dots are the masses of GRB host
galaxies from observations, while the open dots are binned masses of
SDSS galaxies. By comparing the averaged masses of SDSS galaxies
with the those of GRB hosts galaxies, we find that long GRBs prefer
to occur in low mass galaxies. Figure~\ref{metalsfr} presents the
metallicities and SFRs of GRB host galaxies compared to SDSS
galaxies. The gray points represent individual galaxies of SDSS. The
filled dots are the value of GRB host galaxy from observation, while
the open dots are binned SFRs of SDSS galaxies. Obviously, long GRBs
prefer to occur in star-forming galaxies. We also show SFRs and
galaxy masses in Figures~\ref{sfrsdss}. The gray points represent
individual galaxies of SDSS. Filled dots are the value of GRB host
galaxy from observation and open dots are binned SDSS value. By
comparing the averaged SFRs of SDSS galaxies with the those of GRB
hosts galaxies, we find that long GRBs prefer to occur in galaxies
with high SFRs. Figure \ref{avsdss} shows dust extinction $A_V$. We
can see that there is no difference between dust extinction $A_V$
from GRB host galaxies and averaged SDSS value. From these figures,
we can conclude that long GRB host galaxies have low masses, low
metallicities, and high SFRs comparing with the galaxies of SDSS.
This conclusion is consistent with observations, such as Fynbo et
al. (2003), Christensen et al. (2004), Gorosabel et al. (2005),
Fruchter et al. (2006) and Prochaska et al. (2007).

\section{Cumulative redshift distribution of \emph{Swift} long GRBs}

In order to study the cumulative redshift distribution of
\emph{Swift} long GRBs (i.e. the observed duration time is larger
than 2 seconds), we first use long GRBs till GRB 111107A observed by
\emph{Swift}, including dark GRBs from Perley et al. (2009, 2013),
Greiner et al. (2011), and Kr\"{u}hler et al. (2011). This subsample
is also used in Wang (2013). We take the redshift, isotropic energy
$E_{\rm iso}$ and durations of GRBs from Butler et al. (2007, 2010),
and Sakamoto et al. (2011). The luminosity is computed from $L_{\rm
iso}=E_{\rm iso}/[T_{90}/(1+z)]$. The Perley et al. (2009) work
provides us with six redshifts of dark GRBs in our sample. Greiner
et al. (2011) and Kr\"{u}hler et al. (2011) have provided three
additional redshifts of dark GRBs in our sample. Jakobsson et al.
(2012) and Hjorth et al. (2012) obtained 10 new GRB redshifts based
on host galaxy spectroscopy at the ESO Very Large Telescope,
including GRB 050406, GRB 050502B, GRB 050819, GRB 050822, GRB
051001, GRB 051117B, GRB 060719, GRB 070103, GRB 070129, GRB
070419B, which are also including in our sample. We also include 3
three GRBs from Perley et al. (2013). So there are 192 GRBs in our
catalog. We list dark GRBs in Table 2.

The fraction of star formation occurring in galaxies with
metallicities lower than $Z_{\rm crit}$, which can be expressed as
(Stanek et al. 2006)
\begin{equation}\label{psiz}
\Psi(z) = \frac{ \int_{0}^{M_{\rm crit(z)}} \mathrm{SFRM}(M,z)
\phi(M,z) \mathrm{d} M}{ \int_{0}^{\infty} \mathrm{SFRM}(M,z)
\phi(M,z) \mathrm{d} M}
\end{equation}
where $\mathrm{SFRM}(M,z)$ is the star formation rate - stellar mass
relation defined in equation (\ref{eq:SFRz}) and $\phi_(M,z)$ is the
galaxy stellar mass function defined in equation (\ref{eq:Mz}).

\begin{figure}
\centering
\includegraphics[width=0.5\textwidth]{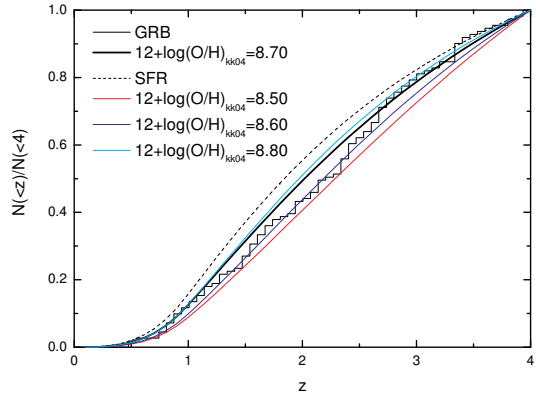} \caption{Cumulative distribution of 111 \emph{Swift} GRBs with
$L_{\rm iso}>10^{51}\rm erg~s^{-1}$ in $z=0-4$ (stepwise solid
line). The dashed line shows the GRB rate inferred from the star
formation history of Hopkins \& Beacom (2006). The solid lines show
the GRB rate inferred from equations (\ref{psiz}) and (\ref{dndz})
using different metallicity cutoffs. } \label{cdf}
\end{figure}

The expected redshift distribution of GRBs is
\begin{equation}\label{dndz}
\frac{d N}{d z}=F_0 \frac{\Psi(z)\dot{\rho}_*(z)}{\langle f_{\rm
beam}\rangle} \frac{dV/dz}{1+z},
\end{equation}
where $F_0$ represents the ability both to detect the trigger of
burst and to obtain the redshift, GRBs that are unobservable due to
beaming are accounted for through $f_{\rm beam}$ and
$\dot{\rho}_*(z)$ is the star formation rate. For the star formation
rate, we use the result of Hopkins \& Beacom (2006), which reads
\begin{equation} \dot{\rho}_*(z)\propto\left\{
\begin{array}{ll}
(1+z)^{3.44},&z\leq0.97,\\
(1+z)^{-0.26},&0.97<z<4.0.
\end{array}\right.\label{sfr1}
\end{equation}
In a flat universe, the comoving volume is calculated by
\begin{eqnarray}
    \frac{dV}{dz} = 4\pi D_{\rm com}^2 \frac{dD_{\rm com}}{d z} \;,
\end{eqnarray}
where the comoving distance is
\begin{eqnarray}
    D_{\rm com}(z) \equiv \frac{c}{H_0} \int_0^z \frac{dz^\prime}{\sqrt{
            \Omega_m(1+z^\prime)^3 + \Omega_\Lambda}} \;.
\end{eqnarray}
In the calculations, we use $\Omega_m=0.27$, $\Omega_\Lambda=0.73$
and $H_0$=71 km~s$^{-1}$~Mpc$^{-1}$ from the \emph{Wilkinson
Microwave Anisotropy Probe} (WMAP) seven-year data (Komatsu et al.
2011). So the number of observed GRBs in the redshift range
$z_1<z<z_2$ is
\begin{equation}
N(z_1,z_2) =  \frac{F_0}{f_{\rm beam}} \int_{z_1}^{z_2} dz\, \Psi(z)
\dot{\rho}_*(z) \frac{dV/dz}{1+z}, \label{Num1}
\end{equation}
where $F_0$ represents the ability both to detect the trigger of
burst and to obtain the redshift. The cumulative distribution of GRB
redshift can be expressed as
\begin{equation}
\frac{N(<z)}{N(<z_{\rm max})}=\frac{N(0,z)}{N(0,z_{\rm max})}.
\end{equation}
Because the observed star formation rate is now reasonably well
measured from $z=0-4$, so we consider GRBs in this redshift range.
In order to avoid the selection effects, we choose the luminosity
cut $L_{\rm iso}> 10^{51}$ erg s$^{-1}$ (Y\"{u}ksel et al. 2008) in
the redshift range $0-4$. We have 111 GRBs in this sub-sample. The
cumulative redshift distribution of these 92 GRBs is shown in Figure
\ref{cdf}. The dashed line shows the GRB rate inferred from the star
formation history of Hopkins \& Beacom (2006). The black line shows
the GRB rate inferred from equations (\ref{psiz}) and (\ref{dndz})
using $Z_{\rm crit}=12+\log\rm(O/H)_{\rm KK04}<8.7$. The cyan, blue
and red lines show the GRB rate inferred from equations (\ref{psiz})
and (\ref{dndz}) using $Z_{\rm crit}=12+\log\rm(O/H)_{\rm
KK04}<8.8$, $Z_{\rm crit}=12+\log\rm(O/H)_{\rm KK04}<8.6$ and
$Z_{\rm crit}=12+\log\rm(O/H)_{\rm KK04}<8.5$, respectively. The
metallicity cutoff about $Z_{\rm crit}=12+\log\rm(O/H)_{\rm
KK04}<8.7$ can well produce the cumulative redshift distribution of
\emph{Swift} long GRBs. We find that the maximum probability occurs
at a metallicity cutoff about $Z_{\rm crit}=12+\log\rm(O/H)_{\rm
KK04}<8.7$ using the Kolmogorov-Smirnov test. Li (2008) also found
that the cumulative redshift distribution of 32 long GRBs can be
well fitted by considering metallicity cutoff $Z\sim 0.3Z_\odot$.
Hao \& Yuan (2013) found that a metallicity cut of $Z\sim
0.6Z_\odot$, which is roughly consistent with our result. The
presence of a host galaxy metallicity ceiling $12+\log\rm(O/H)\leq
8.85$ above which GRBs are suppressed is highly consistent with the
available data is found using the second Swift BAT catalog of GRBs
(Robertson \& Ellis 2012).

\section{Conclusions and discussion}
In this paper, we have investigated the mass distribution of long
gamma-ray burst (GRB) host galaxies and the redshift distribution of
long GRBs. We also compare GRB host galaxies to the galaxies of
SDSS. We calculate the upper limit on the stellar mass of a galaxy
which can produce long GRBs by utilizing the mass-metallicity (M-Z)
relation of galaxies. After comparing with the GRB host galaxies
masses from observation, we find that the observed GRB host galaxy
masses can fit the predicated masses well if GRBs occur in
low-metallicity $12+\log\rm(O/H)_{\rm KK04}<8.7$. GRB host galaxies
have low metallicity, low mass, and high star formation rate
compared with galaxies of seventh data release of the Sloan Digital
Sky Survey. We also study the cumulative redshift distribution of
the latest \emph{Swift} long GRBs by adding dark GRBs and 10 new
GRBs redshifts from TOUGH survey. We find that the observed
discrepancy between the GRB rate history and the star formation
history can be reconciled by considering that GRBs tend to occur in
low-metallicity galaxies with $12+\log\rm(O/H)_{\rm KK04}<8.7$.

We conclude that there is marginal evidence to indicate that GRB
host galaxies are metallically biased tracers of star formation. We
find that the galaxy mass function that includes a smooth decrease
in the efficiency of producing GRBs in galaxies of metallicity above
$12+\log\rm(O/H)_{\rm KK04} = 8.7$ accommodates a majority of the
measured host masses. This is in rough agreement with the
metallicity cutoff $12+\log\rm(O/H)_{\rm KK04}\sim 8.66$ at low
redshift found by Modjaz et al. (2008).

In theory, there are at least three conditions for producing a GRB
with a collapsar (Petrovic et al. 2005; Bromm \& Loeb 2006). First,
in order to form a black hole, the progenitor star must be very
massive. Second, the hydrogen envelope must be lost in order for a
relativistic jet to penetrate through the star. Third, the central
core of progenitor star must have sufficient angular momentum. But a
single massive star has difficulty to fulfill these three
requirements, because of magnetic core-envelope coupling and strong
wind. In order to overcome the two problems, rapidly rotating stars
with low matallcity about $0.1-0.3 Z_\odot$ have been investigated
(Yoon \& Langer 2005; Woosley \& Heger 2006). But some GRB host
galaxies may have much higher metallicity, such as GRB 020819. So
the theory of long GRB formation should be possible to produce GRBs
at high metallicity. Georgy et al. (2012) studied how rotation
modifies the evolution of a given initial mass star towards the
Wolf-Rayet phase and how it impacts the rate of long GRBs. For
solid-body rotation, the explosion of long GRB is restricted to low
metallicity. For internal differential rotation, metallicity also
plays an important role, but long GRB could occur at larger
metallicity, probably at higher than solar metallicity (Georgy et
al. 2012). Non-rotating and rotating star evolutions with
metallicity $z=0.002$ in the mass range from 0.8 to 120 $M_\odot$
had been investigated by Georgy et al. (2013). They also found that
rotation is very important in the stellar evolution. Groh et al.
(2013) investigated the fundamental properties of core-collapse
supernova and GRB progenitors from single stars at solar
metallicity. They found that the GRB progenitors at solar
metallicity have a WO spectral type. Fryer et al. (1999) and
Podsiadlowski et al. (2010) showed that a massive binary can eject
the common envelope, and then produce long GRB at high metallicity.
More recently, numerical simulations of Population (Pop) III star
formation show that they may form in binary or small multiple
systems (Stacy et al. 2010). Pop III stars in binary or small
multiple systems may produce long GRBs efficiently (Bromm \& Loeb
2006).

\section*{Acknowledgements}
We thank the referee for detailed and very constructive suggestions
that have allowed us to improve our manuscript. This work is
supported by the National Basic Research Program of China (973
Program, grant 2014CB845800) and the National Natural Science
Foundation of China (grants 11373022, 11103007, and 11033002).


\begin{deluxetable}{lllllll}\label{hosts}
\tablecolumns{7} \tablewidth{0pc} \tabletypesize{\scriptsize}
\tablecaption{GRB host properties} \tablehead{ \colhead{GRB} &
\colhead{$z$} & \colhead{log(M) [$M_\odot$]} & \colhead{SFR
[$M_\odot$/yr]} & \colhead{12+log(O/H)} & \colhead{$A_V$}&
\colhead{Ref}} \startdata
970228  & 0.695  & 8.65$\pm$0.05 & 0.53       & $8.47$ & 0.63 &1,2,3\\
970508  & 0.835  & $ 8.52\pm0.10$  & 1.14      & ...     & 0.84 &3,4,5 \\
970828  & 0.960   & $ 9.19\pm0.36$    & 0.87      & ...    & 2.13 &5,6,7\\
971214  & 3.420    & $ 9.59\pm0.40$    & 11.40     & ...     & 1.35 &2,5,8\\
980425  & 0.0085   & 9.21$\pm$0.52 & 0.21 & $8.16$ & 1.73  &   5,9,10,11  \\
980613  & 1.097    & $ 8.49\pm0.21$    & 4.70      & ...     & 1.02& 2,12\\
980703  & 0.966    & 9.33$\pm$0.36 & 16.57 & $8.14$ & 1.10 & 3,5,13 \\
990123  & 1.600    & $ 9.42\pm0.49$    & 5.72      & ...     &1.21 &3,5 \\
990506  & 1.310    & $ 9.48\pm0.18$    & 2.50      & ...    & ... &5,14 \\
990705  & 0.842    & $ 10.20\pm0.76$   & 6.96      & ...    & ... &5,9,15,16 \\
990712  & 0.434  & 9.29$\pm$0.02 & 2.39 & $8.10$ & 0.39$\pm0.09  $ &3,17\\
991208 & 0.706     & $ 8.53\pm0.37$    & 4.52      & 8.02     & 0.49&3,18 \\
000210 & 0.846     & $ 9.31\pm0.08$    & 2.28      & ...     & 0.05 &3,19 \\
000418 & 1.118     & $ 9.26\pm0.14$    & 10.35     & ...     & 1.30 & 3,14\\
000911 & 1.058     & $ 9.32\pm0.26$    & 1.57      & ...    & 0.80 &20,21 \\
000926 & 2.036     & $ 9.52\pm0.84$    & 2.28      & ...    & 0.58 &3\\
010222 & 1.480     & $ 8.82\pm0.26$    & 0.34      & ...    & ... &5,22,23,24,25\\
010921  & 0.451  & 9.69$\pm$0.13 & 2.50 & $8.15$ & 1.06$\pm 0.62$&3,25,26  \\
011121  & 0.362  & 9.81$\pm$0.17 & 2.24 & $8.60$ & 0.38  &  27,28   \\
011211  & 2.141  & $ 9.77\pm0.47$    & 4.90      & ...    & ... &29\\
020405  & 0.691  & 9.75$\pm$0.25 & 3.74 & $8.44$ & 1.9$^{+0.6}_{-0.6}$&25,30\\
020813  & 1.255  & $ 8.66\pm1.41$    & 6.76      & ...     & ...&7,25,31 \\
020819 & 0.411  &10.50$\pm$0.14 & 6.86 & $8.98$ & 1.8$\pm0.5        $&32\\
020903  & 0.251  & 8.87$\pm$0.07 & 2.65 & $8.22$ & 0.8$\pm0.2        $&25,33,34\\
021004  & 2.327  & $ 10.20\pm0.18$   & 3.12      & ...    & ... &35,36,37\\
021211  & 1.006  & $ 10.32\pm0.63$   & 3.01      & ...    & 1.78 &25\\
030328  & 1.520  & $ 8.83\pm0.52$    & 3.20      & ...     & 1.06 &38\\
030329  & 0.168  & $ 7.74\pm0.06$    & 0.11      & 7.97     & 0.58 &39,40\\
030528  & 0.782  & 8.82$\pm$0.39 & 15.07 & $8.10$ & ...   & 41,42\\
031203  & 0.1055 & 8.82$\pm$0.43 & 12.68 & $8.02$ & 0.34$\pm 0.05$&43,44\\
040924  & 0.858  & 9.20$\pm$0.37 & 1.88 & $8.23$ & ... &25,45\\
041006  & 0.712  & $ 8.66\pm0.87$ & 0.34      & ...    & ... &25,46\\
050223  & 0.584  & 9.73$\pm$0.36 & 1.44 & $8.66$ & ...&47\\
050826  & 0.296  & 9.79$\pm$0.11 & 9.13 & ... & ...  &48,49\\
051022  & 0.8070 &10.42$\pm$0.18 & 36.46 & $8.65$ & ... & 50,51\\
060218  & 0.0334 & 7.78$\pm$0.08 & 0.05 & $8.13$ & 0.49$\pm0.24 $&52,53,54,55\\
060505  & 0.0889 & 9.41$\pm$0.01 & 0.43 & $8.44         $ & 0.63$\pm0.01$&56\\
060614  & 0.125  & 7.95$\pm$0.13 & 0.01 & $8.24        $ & ...  & 57,58,59,60\\
061126  & 1.159  & $ 10.31\pm0.47$   & 2.38     & ...    & ...& 61\\
061222  & 2.088  & $ 8.04\pm0.21$   & ...      & ...    & ... &62,63\\
070306  & 1.496  & $ 10.36\pm0.21$   & ...      & ...    & 0.13& 63,64,65\\
070714B  & 0.9224  & $ 9.45\pm0.24$   & 0.90 $\pm$0.10   & ...    & ... &66,67,68\\
070802  & 2.455  & $ 9.85\pm0.16$   & ...     & ...    & 0.74& 63,69,70\\
080207  & 2.086  & $ 11.51\pm0.11$   & ...     & ...    & 2.36 &63,71,72\\
080319C  & 1.950  & $12.22\pm0.47$   & ...     & ...    & ... &62,63,69\\
080605  & 1.64  & $ 9.60\pm0.30$   & ...     & ...    & ... &69,73\\
080607  & 3.036  & $ 9.90\pm0.50$   & ...     & ...    & 1.15&63,74 \\
080805  & 1.504  & $ 9.70\pm0.20$   & ...     & ...    & ...&65 \\
081109  & 0.979  & $ 9.82\pm0.09$   & 9.90     & ...    & 1.25 &63,65\\
090205  & 4.650  & $ 10.83\pm0.53$   & ...     & ...    & ... &75\\
090323  & 3.577  & $ 11.20\pm0.75$   & ...     & ...    & ... &76,77\\
090328  & 0.735  & $ 9.82\pm0.08$   & 3.60 $\pm$0.20     & ...    & ... &76,78\\
090417B & 0.345  & $ 9.53\pm0.43$   & ...      & ...    & ... &63,79\\
090926B & 1.240  & $ 10.10\pm0.40$   & ...     & ...    & ...& 65\\
091127  & 0.490  & $ 8.70\pm0.20$   & 0.22 $\pm$0.01     & $8.60$    & ... &80,81\\
100418  & 0.624  & $ 9.28\pm0.28$   & 1.90 $\pm$0.10     & ...    & ... &82,83\\
100621  & 0.542  & $ 8.98\pm0.14$   & ...     & ...    & ...&65,84 \\
120624B  & 2.20  & $ 10.60\pm0.20$   & 23.0     & ...    & 1.0 &85 \\
130427  & 0.34  & $ 9.00\pm0.15$   & 0.90     & 8.70    & 0.05 &86,87 \\
130702  & 0.145  & $ 7.90\pm0.20$   & 0.05     & 8.50    & ... &88,89 \\
\enddata
\tablerefs{ (1)Bloom et al. (2001);(2) Chary et al. 2002; (3)
Christensen et al. 2004; (4) Bloom et al. 1998; (5) Le Floc'h et al.
2006; (6) Djorgovski et al. 2001; (7) Le Floc'h et al. 2003; (8)
Kulkarni et al. 1998; (9) Bloom et al. 2002; (10) Hammer et al.
2006; (11) Sollerman et al. 2005; (12) Djorgovski et al. 2003; (13)
Djorgovski et al. 1998; (14) Bloom et al. 2003; (15) Holland et al.
2000; (16) Le Floc'h et al. 2002; (17) K\"{u}pc\"{u} Yoldas et al.
2006; (18) Castro-Tirado et al. 2001; (19) Piro et al. 2002; (20)
Masetti et al. 2005; (21) Price et al. 2002b; (22) Frail et al.
2002; (23) Fruchter et al. 2001; (24) Galama et al. 2003; (25)
Wainwright et al. 2007; (26) Price et al. 2002a; (27) Garnavich et
al. 2003; (28) K\"{u}pc\"{u} Yoldas et al. 2007; (29) Fynbo et al.
2003; (30) Price et al. 2003; (31) Barth et al. 2003; (32) Jakobsson
et al. 2005; (33) Bersier et al. 2006; (34) Soderberg et al. 2004;
(35) de Ugarte Postigo et al. 2005; (36) Mirabal et al. 2002; (37)
M{\o}ler et al. (2002); (38) Gorosabel et al. 2005a; (39) Gorosabel
et al. 2005b; (40) Th\"{o}ne et al. 2007; (41) Rau et al. 2004; (42)
Rau et al. 2005; (43) Cobb et al. 2004; (44) Prochaska et al. 2004;
(45) Wiersema et al. 2008; (46) Soderberg et al. 2006a; (47)
Pellizza et al. 2006; (48) Mirabal et al. 2007; (49) Ovaldsen et al.
2007; (50) Castro-Tirado et al. 2007; (51) Rol et al. 2007; (52)
Cobb et al. 2006a; (53) Pian et al. 2006; (54) Sollerman et al.
2006; (55)Wiersema et al. 2007; (56) Th\"{o}ne et al. 2008; (57)
Cobb et al. 2006b; (58) Della Valle et al. 2006; (59) Gal-Yam et al.
2006; (60) Mangano et al. 2007; (61) Perley et al. 2008; (62) Perley
et al. 2009; (63) Perley et al. 2013; (64) Jaunsen et al. 2008; (65)
Kr\"{u}hler et al. 2011; (66) Cenko et al. 2008; (67) Graham et al.
2009; (68) Fong \& Berger 2013; (69) Fynbo et al. 2009; (70)
Milvang-Jensen et al. 2012; (71) Hunt et al. 2011; (72) Svensson et
al. 2012; (73) Kr\"{u}hler et al. 2012; (74) Wang et al. 2012; (75)
D'Avanzo et al. 2010; (76) McBreen et al. 2010; (77) Savaglio et al.
2012; (78) Cenko et al. 2011; (79) Holland et al. 2010; (80) Berger
et al. 2011; (81) Vergani et al. 2011; (82) Niino et al. 2011; (83)
de Ugarte Postigo et al. 2012; (84) Greiner et al. 2013; (85) de
Ugarte Postigo et al. 2013; (86) Xu et al. 2013; (87) Perley et al.
2014; (88) Kelly et al. 2013; (89) Singer et al. 2013.}
\end{deluxetable}


\begin{deluxetable}{lllll} \label{grbs}
\tablecolumns{5} \tablewidth{0pc} \tabletypesize{\scriptsize}
\tablecaption{GRB Catalog} \tablehead{ \colhead{GRB} & \colhead{$z$}
& \colhead{$E_{\rm iso}$ [$10^{52}$ erg]} & \colhead{$T_{90}$ [s]} &
\colhead{$L_{\rm iso}$ [$10^{52}$ erg s$^{-1}$]} } \startdata
\multicolumn{5}{l}{Dark GRBs} \\
\hline
050915A   & 2.53  &$1.94_{-0.6}^{+2.6}$     & 53.4 &$0.13_{-0.04}^{+0.18}$  \\
060210 & 3.91      &$ 42.0_{-8.0 }^{+35.0 }$   & 242   &$ 0.85_{-0.16 }^{+0.71 }$     \\
060510B    & 4.94      &$ 23.0_{-4.0 }^{+10.0 }$   & 263   &$ 0.52_{-0.09 }^{+0.23 }$     \\
061222A    & 2.09      &$ 67.4_{-12.8}^{+35.3}$  & 96.0  &$ 2.17_{-0.41 }^{+1.14 }$     \\
070521    & 1.35&$ 25.2_{-8.8}^{+22.0}$ & 38.6   &$ 1.54_{-0.53 }^{+1.34 }$\\
081109   & 0.98   &$ 4.1_{-2.2}^{+2.6}$    & 221   &$ 0.037_{-0.020 }^{+0.023 }$\\
080319C   & 1.95  &$ 6.0_{-1.0 }^{+5.0 }$     & 29.5  &$ 0.60_{-0.10 }^{+0.50 }$\\
080516    & 3.60   &$ 12.0_{-4.8}^{+6.0}$   & 5.75  &$ 9.57_{-3.83 }^{+4.79 }$\\
081228    & 3.40   &$ 3.7_{-1.3 }^{+1.6 }$  & 3.00  &$ 5.36_{-1.84 }^{+2.30 }$\\
051008    & 2.90   &$ 9.6_{-1.0 }^{+1.5 }$  & 16.0  &$ 2.34_{-1.25 }^{+2.10 }$\\
090404    & 3.00   &$ 5.9_{-2.3 }^{+1.8 }$  & 84.0  &$ 0.29_{-2.54 }^{+2.42 }$\\
090709A    & 1.80   &$ 20.9_{-2.5 }^{+3.6 }$  & 89.0  &$ 0.66_{-2.83 }^{+3.35 }$\\
\hline
\multicolumn{5}{l}{TOUGH GRBs} \\
\hline
050406  & 2.7   &$0.23_{-0.08}^{+0.20}$  & 5.0  &$ 0.17_{-0.06}^{+0.15}$\\
050502B  &5.2   &$3.84_{-0.93}^{+7.61}$  & 17.4  &$ 1.37_{-0.33}^{+2.71}$\\
050819  & 2.5   &$1.02_{-0.34}^{+1.48}$  & 47  &$ 0.076_{-0.0025}^{+0.11}$\\
050822  & 1.434   &$2.55_{-0.26}^{+3.15}$  &105  &$ 0.059_{-0.006}^{+0.073}$\\
051001  & 2.43   &$2.09_{-0.29}^{+0.56}$  & 56  &$ 0.128_{-0.018}^{+0.034}$\\
051117B  & 0.481   &$0.018_{-0.005}^{+0.02}$  &10.5  &$ 0.0026_{-0.0007}^{+0.0029}$\\
060719  & 1.532   &$1.60_{-0.25}^{+1.85}$  & 57  &$ 0.071_{-0.011}^{+0.082}$\\
070103  & 2.62   &$0.58_{-0.14}^{+0.59}$  & 10.9  &$ 0.191_{-0.046}^{+0.195}$\\
070129  & 2.338   &$2.89_{-0.55}^{+1.24}$  & 92  &$ 0.105_{-0.02}^{+0.045}$\\
070419B  & 1.959   &$15.7_{-2.94}^{+10.8}$  & 134  &$ 0.346_{-0.065}^{+0.238}$\\
\enddata
\end{deluxetable}

\begin{figure}\centering
\includegraphics[width=0.5\textwidth]{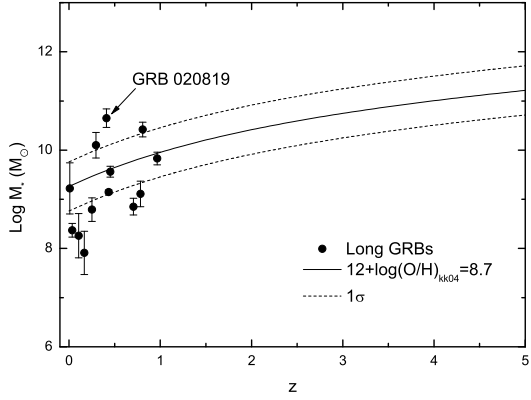} \caption{GRB host galaxy mass distribution.
The host galaxy masses are taken from Levesque et al. (2010a). The
solid line represents the upper limits of the stellar mass of a GRB
host galaxy given a metallicity cutoff of $12+\log\rm(O/H)_{\rm
KK04}=8.7$. The dashed lines represent the scatter in the upper
limit imposed by the 1$\sigma$ scatter of the M-Z relation.}
\label{Leva}
\end{figure}

\end{document}